\documentclass{article}
\usepackage{spconf,amsmath,graphicx, amsfonts,booktabs}
\usepackage{enumitem}
\setlist{nosep, leftmargin=14pt}

\usepackage{mwe} % to get dummy images

\title{NUDF: Neural Unsigned Distance Fields for high resolution 3D medical image segmentation}
\name{Kristine S{\o}rensen$^{\star}$ \quad Oscar Camara$^{\ddag}$ \quad Ole de Backer$^{\dagger}$ \quad Klaus F. Kofoed$^{\dagger}$  \quad Rasmus R. Paulsen$^{\star}$}

\address{$^{\star}$Department of Applied Mathematics and Computer Science\\Technical University of Denmark, Kgs. Lyngby, Denmark \\
    $^{\ddag}$BCN MedTech, Universitat Pompeu Fabra, Barcelona, Spain \\
     $^{\dagger}$The Heart Center, Rigshospitalet, University of Copenhagen, Copenhagen, Denmark}

\begin{document}
%\ninept
%
\maketitle
\begin{abstract}
%Include abstract text
Medical image segmentation is often considered as the task of labelling each pixel or voxel as being inside or outside a given anatomy.
Processing the images at their original size and resolution often result in insuperable memory requirements, but downsampling the images leads to a loss of important details. 
Instead of aiming to represent a smooth and continuous surface in a binary voxel-grid, we propose to learn a Neural Unsigned Distance Field (NUDF) directly from the image. 
The small memory requirements of NUDF allow for high resolution processing, while the continuous nature of the distance field allows us to create high resolution 3D mesh models of shapes of any topology (i.e. open surfaces). 
We evaluate our method on the task of left atrial appendage (LAA) segmentation from Computed Tomography (CT) images. 
The LAA is a complex and highly variable shape, being thus difficult to represent with traditional segmentation methods using discrete labelmaps. 
With our proposed method, we are able to predict 3D mesh models that capture the details of the LAA and achieve accuracy in the order of the voxel spacing in the CT images.

%In this paper, we propose to use neural implicit distance fields to represent the anatomy of interest in a segmentation algorithm. 
%The continuous nature of the distance field allows us to create high resolution 3D mesh models from the input image and can represent shapes of any topology (ie. open surfaces). 
%We evaluate our method on the task of Left Atrial Appendage (LAA) segmentation from Computed Tomography (CT) images. 
%The LAA is a complex and highly variable shape and is thus difficult to represent with traditional segmentation methods using binary labelmaps. 
%With our proposed method we were able to predict 3D mesh models that captured the details of the LAA and achieved accuracies in the order of the voxel spacing in the CT images.
%Code available at \textit{Provided at publication}.

%In this paper we propose a novel method for creating detailed 3D mesh models from medical images without excessive memory requirements. 
%Our model extracts multi-scale feature maps from the image using a combination of 3D convolutions and max-poolings. 
%The feature maps are sampled at a continuous location and fed into a decoder network predicting the distance value at the given location. 

\end{abstract}
\begin{keywords}
Unsigned distance fields, image segmentation, mesh modelling, left atrial appendage, computed tomography
\end{keywords}
\section{Introduction}
\label{sec:intro}
3D medical imaging techniques (such as computed tomography (CT)) are becoming increasingly available for creating high resolution images of human anatomies. 
Today many of these images are manually analyzed and annotated, but the fast advance in deep learning (and especially Convolutional Neural Networks (CNNs)) are making automatic methods more feasible. 
Volumetric CNNs dominate the field of 3D medical image segmentation, but suffer from high memory requirements, which force us to downsample the input images and thereby risk loosing important details. 
Multiple tactics has been explored to mitigate this, including slicewise or patchwise processing as well as substituting the 3D convolutions with more lightweight options~\cite{Jin2018,Sundgaard2020,zhao2021}. 

In clinical and research practice, we often require a mesh representation for visualization, simulation and alike. 
When using volumetric CNNs the output is usually a voxel volume, which can be converted into a mesh model using isosurfacing, smoothing and decimation algorithms. 
%The post processing steps do however increase the risk of introducing artefacts or removing small details. 
Alternatively 3D mesh models can be generated directly from the image input with a combination of 3D convolutions for encoding and Graph Convolutions for decoding~\cite{Wickramasinghe2020,Wickramasinghe2021}. 

In the computer vision community there is a current interest in representing shapes as deep neural implicit functions such as for example signed distance fields (SDFs)~\cite{Park2019}, unsigned distance fields (UDFs)~\cite{Chibane2020a,juhl2021implicit} or occupancy functions~\cite{Chibane2020b}. 
Within image segmentation is has been shown that an SDF can be predicted directly from an image~\cite{Juhl2019}, but their method discretized the SDF to the image grid, whereas our proposed NUDF learns a continuous function. 
%\cite{Raju2021} and \cite{Barrowclough2021} respectively learns an implicitly defined shape model and a spline representation from images. 
Other methods aim to learn an implicit shape model~\cite{Raju2021} or a spline representation~\cite{Barrowclough2021} from an image.
Compared to this work, our NUDF are not constrained to fit within a shape model or has to be represented by stacks of 2D splines. 

\begin{figure*}[tb]
\begin{minipage}[b]{1.0\linewidth}
  \centering
  \centerline{\includegraphics[width=\linewidth]{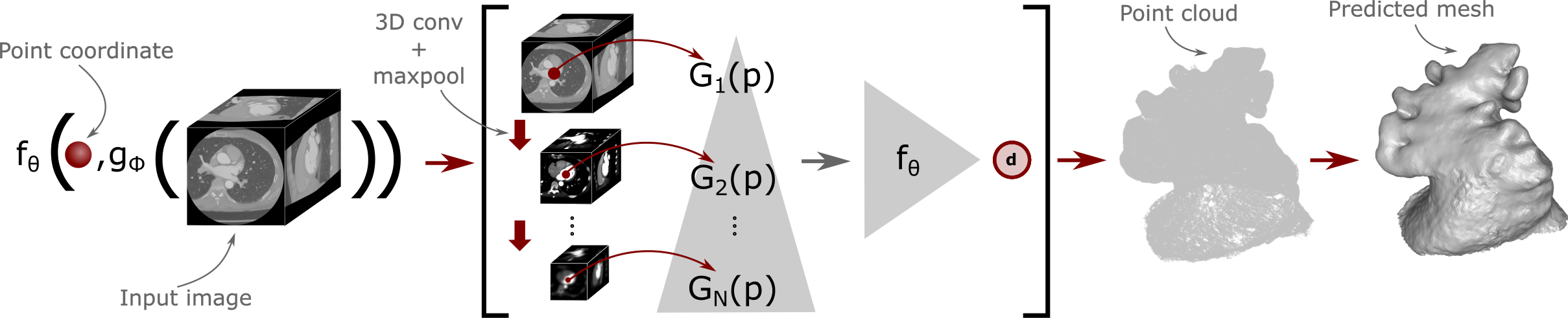}}
\end{minipage}
\caption{Overview of the proposed method for creating 3D mesh models from high resolution input images. The input image is processed through a series of 3D convolutions and maxpoolings to produce feature maps $G_{1..N}$. The feature maps are sampled at continuous point coordinates (red points) and inputted to a fully connected neural network (FCNN) predicting the distance from the point to the surface. By repeating this for many points we can learn a distance field representing the anatomy of interest, from which a dense point cloud and/or a triangulated mesh can be extracted.}
\label{fig:overview}
\end{figure*}

In the following paragraphs we will present our novel NUDF segmentation scheme.
The scheme is illustrated in Figure \ref{fig:overview} and is based on a convolutional image encoder and a point-wise distance field decoder approximated by a fully connected neural network (FCNN).
A unique sampling scheme was developed to focus the training on complex parts of the surface. 
Finally, the predicted UDFs were converted to 3D mesh models in high resolution\footnote{Code: \texttt{https://github.com/kristineaajuhl/NUDF}}.

%In this paper we propose a novel high resolution 3D medical image segmentation network based on neural unsigned distance fields. 
%The network reduces the computational burden by substituting the memory-heavy decoder path from a typical CNN with a more lightweight fully connected neural network (FCNN). 
%We extract features of different resolutions from the image using 3D convolutions and maxpoolings and sample these multi-resolution feature maps continuously. 
%The sampled features serve as an input to the FCNN, which outputs the unsigned distance from the sample point to the nearest point on the surface. 
%We develop and describe a unique sampling scheme that allows us to focus the training at complex parts of the surface and finally describe how the predicted UDFs can be converted to 3D mesh models in high resolution. 

%\section{Related work}
%\label{sec:sota}
%\input{tex/sota}

\section{Data}
\label{sec:data}
We evaluated our method on the task of segmentation and mesh model creation of the Left Atrial Appendage (LAA) from CT images. 
The LAA is a complex pouch on the Left Atrium (LA) and its morphology is known to correlate with the stroke risk for patients with atrial fibrillation~\cite{Glikson2020}. 
%The medical doctors rely on mesh models of the individual anatomies to investigate this correlation and to choose the correct treatment plan for patient with high stroke risk. 

Our data set consists of 106 randomly selected healthy participants, who have undergone a CT examination at Rigshospitalet, Copenhagen in the period 2010-2013 for research purposes.
%Participant consent are obtained for all analyzed cases and the data set is fully anonymized. 
The LA and LAA are manually segmented in 3D Slicer and converted into mesh models. The LAA is separated from the LA with a plane at the ostium using an automatic method aiming to slice the LAA neck at its narrowest point. 
%This results in a open triangulated surface representing the LAA.

\section{Methods}
\label{sec:methods}
\subsection{Neural Implicit Function learning}
A UDF is an implicit continuous function that, for any point in space $\textbf{p}$, outputs the unsigned distance $d$ from $\textbf{p}$ to the closest point on the surface. 
The UDF can be defined as $\text{UDF}(\textbf{p}) = d : \textbf{p} \in \mathbb{R}^3, d \in  \mathbb{R}^+$. 
The underlying surface can be extracted as the isosurface at $\text{UDF}(\cdot) = 0$. 

We propose to learn a neural representation of such distance function using a deep neural network (Figure~\ref{fig:overview}):

\begin{equation}
    f_\Theta (\textbf{p},g_{\Phi}(\textbf{X})) \approx \text{DF}(\textbf{p}),
\end{equation}

\noindent where $g_{\Phi}(\textbf{X})$ denotes an encoding of the input image $\textbf{X}$ by the network $g_{\Phi}$ and $f_\Theta$ a FCNN approximating the UDF. 
The encoder $g_{\Phi}$ consists of a series of 3D convolutions and maxpoolings producing feature maps $\textbf{G}_1,...,\textbf{G}_N$ of decreasing resolution. 
Instead of decoding the information in a symmetric convolutional decoding path, we instead regress the distance field at a continuous query point $\textbf{p}$.
To incorporate neighborhood information, the feature maps are sampled at the query point and at surrounding points at a distance $l$ along the Cartesian axes and fed into $f_\Theta$ regressing the unsigned distance.

We trained the image encoder $g_{\Phi}$ and the point-wise decoder $f_\Theta$ jointly. 
For each example $i$ in all training examples $\Omega$, we prepared a set consisting of an input image $\textbf{X}_i$, a collection of $j \in N$ points $\textbf{P}_{i,j}$ and the distance from each point to the surface $\textbf{d}_{i,j}$. 
The network was optimized using L1-loss between the predicted and ground truth distances for a collection of points as follows:

\begin{equation}
    \mathcal{L}_1 = \sum_{i\in \Omega} \sum_{j \in N} | f_\Theta (\textbf{p}_{i,j},g_{\Phi}(\textbf{X}_i))-\text{d}_{i,j} |.
\end{equation}

\subsection{Sampling strategy}\label{sec:sampling} 
\begin{figure}[htbp]
    \centering
    \includegraphics[width=0.49\textwidth]{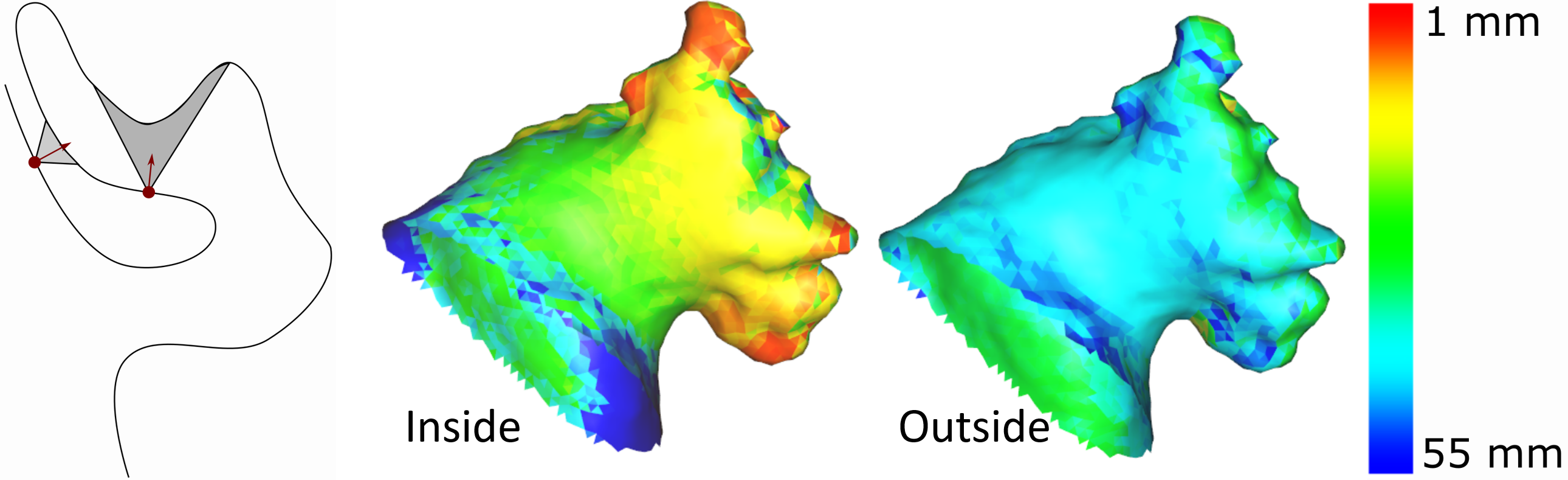}
    \caption{\textbf{Left)} A 2D illustration of the shape diameter (SD) calculation, where a cone of rays (grey area) is cast from each point in the direction of the normal (red arrow). \textbf{Right)} Example of the SD evaluated with the original and with flipped normals.}
    \label{fig:SD}
\end{figure}

%The simplest sampling strategy involves sampling a set of points $\textbf{P}_i$ on the surface and perturbing the point coordinates with Gaussian noise $\mathcal{N}(0,\sigma^2)$, where $\sigma^2$ is a user-defined variance. 
%We however observed two problems with this strategy: 1) all parts of the surface are sampled equally with no regards to the surface complexity and 2) there is few samples inside thin structures or between almost touching structures.
%points located at parts of the surface that are spatially close to other parts of the surface (ie. thin or almost-touching parts) have few samples in this space as the gaussian perturbation make the majority of the points traverse through the shape.
%\textbf{XXX: Consider making figure about this.}
%We denote such surface parts as "volatile structures" and the space between the surfaces as "volatile gaps". 
%To mitigate these problems, we define a sampling scheme that are dependent on the Shape Diameter (SD) \cite{Kovacic2010}. 
When training the algorithm, it is important to have many UDF samples in the vicinity of the surface to capture the fine details.
Furthermore, it is important to have many samples in between surface parts that are located close to each other (i.e. \textit{thin} or \textit{almost-touching} structures). 
To achieve this, we define a sampling scheme based on the Shape Diameter (SD) of the shape. 
As visualized on Figure \ref{fig:SD}, the SD is calculated by casting a cone of rays in the direction of the normal in each point on the surface and measuring the average distance of all rays' nearest intersection with the mesh.  

We sampled points randomly on the surface and perturbed them $\mathcal{N}(0,\text{SD}/4)$ along their barycentric normal.
The process was repeated with flipped normals to account for both \textit{thin} and \textit{almost-touching} structures, as well as sampling of open surfaces. 
To control the density of the sampling across different regions, we introduced the sampling factor $\lambda$, sampling points $\lambda$ times more densely in areas with a small SD. 

\subsection{Meshing unsigned distance fields}
Finding the isosurface at $\text{UDF}(\cdot) = 0$ is challenged by the fact that the UDF never crosses zero (but gets infinitely close) and that the UDF is undifferentiable at zero. 
We make use of the method from~\cite{Chibane2020a} to create a dense point cloud from the neural UDF. 
Initially, a set of points $\textbf{p}$ were sampled uniformly within the image volume, the distance $d =  f_\Theta (\textbf{p},g_{\Phi}(\textbf{X}))$ was predicted and the gradient of the distance field $\nabla_\textbf{p}  f_\Theta (\textbf{p},g_{\Phi}(\textbf{X}))$ computed using standard backpropagation. 
For a true distance field, the norm of the gradient equals one ($||\nabla_\textbf{p}  f_\Theta (\textbf{p},g_{\Phi}(\textbf{X}))|| = 1$), we could theoretically project any point $\textbf{p}$ directly to the surface as follows: 

\begin{equation}\label{eq:project}
    \textbf{q} = \textbf{p} -  f_\Theta (\textbf{p},g_{\Phi}(\textbf{X})) \cdot \nabla_\textbf{p}  f_\Theta (\textbf{p},g_{\Phi}(\textbf{X})).
\end{equation}

Since $f_\Theta (\textbf{p},g_{\Phi}(\textbf{X}))$ is an approximation of the real UDF, Equation \ref{eq:project} does not hold in practice. 
Good results can however be obtained by repeating the process for multiple iterations each time predicting a new distance and gradient. 
In many cases the output from a segmentation is used for downstream tasks such as visualization or simulation, which requires high quality triangulated meshes. 
Such meshes are difficult to generate from a point cloud with noise and varying sampling density. 
We found that subsampling the point cloud from approximately $10^5$ to $10^3$ points using Poisson Disk Sampling~\cite{Corsini2012} before reconstructing the surface with a Screened Poisson Reconstruction~\cite{Kazhdan2013} could handle such issues. 
Since Poisson Reconstruction aims to create closed surfaces, we removed all triangles in the reconstructed mesh that are not supported by any point in the original point cloud within 0.35 mm. 
Finally, all holes in the surface with an approximated radius smaller than 3.5 mm were closed and the largest connected component extracted.
%The meshing is carried out in Python using MeshLab \cite{meshlab} and VTK \cite{VTK}. 

\subsection{Implementation details}
\noindent\textbf{Region-of-interest detection}
A region-of-interest (ROI) around the LAA was predicted using a 3D U-net~\cite{Ronneberger2015}. 
The LAA was converted to a binary labelmap by closing the ostium hole and the ROI network was trained at a resolution of $64^3$ on the same training set as the remaining of the algorithm. 
The size of the ROI was 70 mm on each side and it was placed at the center-of-mass of the predicted labelmap. 
Since any off-the-shelf method could be used for detecting the ROI, we refer the reader to the code for further details.

\noindent\textbf{Dataset preparation.}
The ROI images were clipped at $\pm1000$ HU and normalized to $[-1;1]$.
We normalized the image coordinates to lie between [-1,1] and scaled the isosurface accordingly. 
For each training example we sampled 10 000 points uniformly across the entire image volume and 100 000 close to the surface.
The SD was computed using an GPU accelerated implementation in MeshLab~\cite{meshlab} with a cone amplitude of 45 degrees. 
Areas with $\text{SD} < 10$ mm were sampled $\lambda = 500$ times more densely than other regions, and areas with $\text{SD} > 10$ mm were sampled with standard deviation $\sigma_1 = 10$ mm and $\sigma_2 = 20$ mm. 
We measured the unsigned distance from each of the sampled points to the the surface using an octree based spatial search algorithm as implemented in VTK~\cite{VTK}.

\noindent\textbf{Network architecture.}
Both the encoding and decoding networks were chosen similar to the networks from~\cite{Chibane2020b}.
The image encoder $g_\Phi$ consists of 3D convolutions, batch normalization and 3D maxpoolings. 
We experimented with input resolutions of $64^3$ and $128^3$. 
The decoding network $f_\theta$ is a FCNN consisting of two hidden layers with each 256 units. 
The number of inputs to $f_\theta$ is dependent on the image input size and the number of feature channels in each convolution. 

\noindent\textbf{Training details.}
We split the dataset into 76 images for training, 10 images for validation and 20 images for testing. 
To avoid overfitting, we augmented the training data online with a 50\% probability for $\pm 20$ voxels translation, $\pm 10^{\circ}$ rotation and Gaussian noise on the normalized image intensities with $\mathcal{N}(0,0.05)$.
We also used $10\%$ dropout in the encoder and made use of early stopping. 
The network was trained and tested on a single Nvidia Titan X with 12GB GPU RAM using a batch size of 4 for all models.

\section{Experiments}
\label{sec:results}
We compare the results from our proposed neural implicit distance prediction to a standard 3D U-net \cite{Ronneberger2015}. 
The encoding path of the U-net is chosen similar to the encoding network of our model and the decoder is constructed symmetrically.
The U-net takes $64^3$ ROI images as input and outputs a probability map in the same dimensions. 
The probability map is trilinearly upsampled to the original resolution and isosurfaced using marching cubes \cite{lorensen1987}. 

%\begin{table}[tbh]
%\begin{tabular}{lccccccc} \toprule
%                   & \textbf{Image Size}  & \textbf{Mean CD} & \textbf{Median CD} & \textbf{Mesh Acc.} & \textbf{Mesh Compl.} \\ \midrule
%\textbf{U-net}     & 64                     &      0.877       &     0.289    & 2.712             &  94.71                   \\
%\textbf{SDF U-net} & 64                   &                     &                &              &                &                        \\
%\textbf{Ours}      & 64                      &     0.425      &      0.179    & 0.905          &     98.16                          \\
%\textbf{Ours}      & 128                   &       0.453      &      0.185    &  1.054           & 98.26                         \\ \bottomrule      
%\end{tabular}
%\label{tab:results}
%\caption{Table caption}
%\end{table}

\begin{table}[tbh]
\begin{tabular}{lccccccc} \toprule
                   & \textbf{Input}  & \textbf{Chamfer}  & \textbf{Mesh} & \textbf{Mesh} \\ 
                   & \textbf{size}  & \textbf{Distance}  & \textbf{Accuracy} & \textbf{Completion} \\ \midrule
\textbf{U-net}     & 64                     &      0.877 mm          & 2.712 mm            &  94.71 \%                 \\
%\textbf{SDF U-net} & 64                   &                     &                &              &                &                        \\
\textbf{NUDF}      & 64                      &     0.425 mm         & 0.905 mm         &     98.16 \%                     \\
\textbf{NUDF}      & 128                   &       0.453 mm         &  1.054 mm       & 98.26 \%                         \\ \bottomrule      
\end{tabular}
\label{tab:results}
\caption{Qualitative evaluation results comparing NUDF at different resolutions to a 3D U-net}
\end{table}

\begin{figure*}[htb]
\begin{minipage}[b]{0.92\linewidth}
  \includegraphics[width=\linewidth]{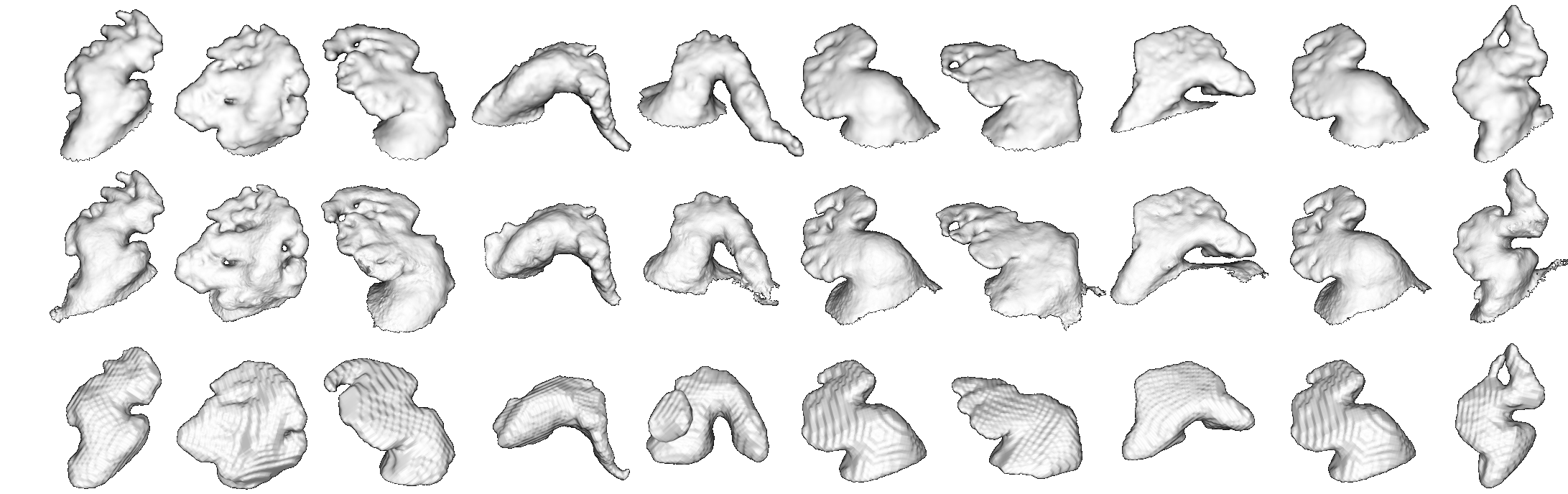}
\end{minipage}
\caption{Examples of meshes from manual segmentation, NUDF and a standard 3D U-net. The examples are ordered from left to right as best to worst chamfer distance on our proposed method.}
\label{fig:results}
\end{figure*}

%We compare the methods quantitatively to the manual annotation using three metrics. 
%Symmetric Chamfer Distance (CD) measures the distance from all points in the manual mesh to the predicted mesh and vice versa, Mesh Accuracy (MA) calculates the distance from the manual segmentation within which 90\% of the predicted points lies within and Mesh Completion (MC) measures the percentage of the manual segmentation that are covered within 2 mm. 
We evaluate our method using symmetric Chamfer Distance, Mesh Accuracy at $90\%$ and Mesh completion at 2 mm as seen in Table \ref{tab:results}. 
It is evident that the proposed method performs better than the 3D U-net in all measurements despite using the same resolution. 
It is worth noting that the U-net segmentations are closed surfaces, which means the Chamfer Distance and Mesh Accuracy are slightly overestimated, as large errors will be measured at the artificial closing. 
The Mesh Completion is however unaffected by this, and still reveals a significant improvement using our method. 
Since we are using a lighter decoding path, we are able to increase the input resolution to $128^3$, but did not observe any difference in performance. 
It was not possible to train the 3D U-net at higher resolutions due to memory constraints. 

Figure \ref{fig:results} shows 10 test examples ranging from the lowest to the highest Chamfer Distance sorted on the results from our method. 
It is evident that our method is able to recover the shape details of the highly complex LAAs, whereas the 3D U-net only is able to recover the overall LAA shape.
The low resolution of the probability map also creates voxel-artefacts. 
These can be removed by smoothing the surface, but we thereby risk removing more details from the models.

\section{Discussion and conclusion}
\label{sec:discussion}
% Comparison to other LAA segmentation algorithms - no on else creates a mesh - important for fluid simulations, device selection etc.

% Resolution vs training/test-time. 

% Predictions - can be parallelized.

% Importance of open surface models - models with subvoxel accuracy

% Difficulties with meshing algorithms -> holes in 

The best predictions in our test set (Figure \ref{fig:results} (Left)) are almost perfect replicates of the manual segmentation.
As the performance declines (moving to the right), we see that the main cause of error occurs when detecting where to end the open surface. 
This is expected, since there is no direct cues in the image, on where to place the cut between the LA and the LAA. 
Besides errors near the ostium, we observe that our model sometimes have additional details, that are not on the manual segmentation. 
Visual inspection of these cases overlayed on the image revealed that our predicted mesh often follows the image better than the manual segmentation.
When creating meshes from the manual image segmentation, we smoothed the meshes slightly to remove voxel-artefacts, which may explain this observation. 

The resolution of the original images are 0.429 mm/voxel, while we process the ROI with 1.094 mm/voxel at input size $64^3$.
Despite the downsampling of the images, we still achieve chamfer distances in the same order of magnitude as the original voxel spacing.
We are able to train models with input images of up to $512^3$ voxels without exceeding 12 GB memory, which means we could train with the full resolution images without first detecting the ROI. 
Increasing the input-size to $512^3$ however increased the expected training time from $\approx2$ days at $128^3$ to an estimated $\approx100$ days at $512^3$ assuming similar convergence at about 1000 epochs. 
For this reason we opted for the two-step approach.

State-of-the-art LAA segmentation algorithms \cite{Jin2018,Leventic2019} use traditional binary maps to represent the LAA, which makes it difficult to compare their DICE-scores to our mesh-based evaluations. 
Compared to their methods, NUDF are capable of handling open surfaces and surfaces with details smaller than the voxel resolution.
We did not compare to other methods predicting the mesh such as Voxel2Mesh \cite{Wickramasinghe2020}, since they require 32 GB memory to train the model or a downsampling of the mesh, so severe that all details from the LAA is lost.

We attribute the success of our NUDF approach largely to the ability to decode continuous DFs and thereby represent high resolution segmentations independent of the size of the input image. 
The downside of this is, that multiple passes (forward and backwards) in the network are required to obtain the dense point cloud needed for triangulation. 
In comparison a regular 3D U-net requires only one forward pass, but it takes time to do transposed convolutions and upsample the low resolution probability map. 
In practice the time-difference of the two approaches are therefore marginal. 

%One of the main challenges from the proposed method is the meshing of the UDF.
%The neural implicit distance field can easily be converted to a dense point cloud, but it showed difficult to obtain high quality surfaces from such points.
%An important parameter in this is the distance used to remove parts of the surface from the Poisson reconstruction that are not supported by the point cloud. 
%Selecting this distance too small may cause holes in the mesh at the less densely sampled areas, whereas a too large distance results in inclusion of surface that are unsupported interpolations. 

In conclusion, we argue that the proposed NUDF is a novel method for learning high resolution segmentations of 3D medical images. 
The method creates smooth, continuous distance fields that can be converted into triangulated meshes useful for downstream tasks such as visualization, fluid simulations, etc..
The network requires limited memory capacity, but is able to predict detailed meshes with errors in the same order of magnitude as the voxel spacing in the original image. Finally, NUDF can also be used for disjoint open surfaces.

\section{Compliance with ethical standards}
\label{sec:ethical}
Participation was conducted following the declaration of Helsinki and approved by the ethical committee (H-KF-01-144/01).
%The Danish National Committee on Biomedical Research Ethics approved the research protocol (H-KF-01-144/01), and all patients gave oral and written consent.
%The research protocol (\textit{Anonymous number}) are approved by \textit{Anonymous Comittee} and all patients gave oral and written consent.
 The data is fully anonymized including removal of all patient specific information except for the CT image data. 

\section{Acknowledgements}
\label{sec:ack}
This work was supported by a PhD grant from the Technical University of Denmark - Department of Applied Mathematics and Computer Science (DTU Compute) and the Spanish Ministry of Science, Innovation and Universities under the Retos I+D Programme (RTI2018-101193-B-I00).
%This work is supported by \textit{Anonymous grant 1} and \textit{Anonymous grant 2}.

\bibliographystyle{IEEEbib}
\bibliography{refs}

\end{document}